# Integrated quantum-secured cryptographic memory in customizable single nanodiamonds

Tongtong Zhang,[1]† Jiaqi Li, [1]† and Zhiqin Chu[1, 2, *]

[1] Department of Electrical and Computer Engineering, The University of Hong Kong, Pokfulam Road, Hong Kong, China
[2] School of Biomedical Engineering, The University of Hong Kong, Hong Kong, China
† These authors contributed equally to this work.
* Corresponding author: Z.Q.C. (zqchu@eee.hku.hk)

## Abstract

As advanced physical cloning and cyber-extraction techniques proliferate, achieving absolute hardware-level information security has become a paramount global challenge. Conventional architectures systems are fundamentally vulnerable because deterministic memory and cryptographic hardware are physically decoupled, exposing sensitive data to interception. While physical unclonable functions (PUFs) offer robust authentication, integrating them with high-density data writing within a unified nanoscale medium remains elusive. Here, we present an intrinsically secure all-in-one architecture that inextricably fuses high-dimensional data encoding with multiscale physical encryption within single nanodiamonds (NDs). Through precise focused-electron-beam irradiation, we achieve single-nanoparticle defect engineering, enabling deterministic base-36 data writing within individual NDs. Crucially, this customized memory layer is permanently locked within the medium's inherent stochasticity, combining macroscopic spatial PUFs with strictly irreproducible atomic-scale quantum fingerprints extracted via optically detected magnetic resonance (ODMR). By seamlessly embedding deterministic memory into an unclonable physical environment, our approach intrinsically eliminates the vulnerabilities of separated hardware, providing an unconditionally secure solid-state foundation for tamper-proof memory chips, zero-trust networks and next-generation data vaults.

## Introduction

In an increasingly digitized society, the security of sensitive information has become a cornerstone of global economic and geopolitical stability. However, conventional hardware architectures face a fundamental and escalating threat: the physical decoupling of data storage and cryptographic components (*1-3*). In current paradigms, deterministic information is stored in standard memory modules (*4, 5*), while security is provisioned by separate cryptographic processors or physical unclonable functions (PUFs) (*6-8*). This inherent segregation creates a critical security gap, leaving data vulnerable to sophisticated physical extraction, cloning, and side-channel attacks during the transfer between storage and encryption layers (*3, 9*). To achieve hardware-intrinsic security, the next generation of information technology demands an all-in-on architecture, a unified physical medium where high-capacity data writing is inextricably fused with intrinsically unclonable encryption.

Considerable efforts have been made to bridge the gap between high-density storage and physical security (**Table S1**). For instance, biomolecular media such as synthetic DNA (*10-12*) offer immense storage density and steganographic potential, yet they suffer from slow read-write speeds and poor compatibility with existing solid-state electronics. Alternatively, emerging non-volatile memories like resistive random-access memory (ReRAM) (*13-15*) leverage stochastic conductive path formation as a source of PUF entropy, but they are often limited by low encoding dimensions and susceptibility to environmental noise and read instability. Optical data storage, particularly in robust transparent media like glass (*16*) or bulk diamond (*17,*

*18*), has demonstrated extraordinary longevity (*19*) and three-dimensional (3D) capacity (*20*). Despite their impressive capacities, these optical storage approaches are limited by their deterministic nature. Because the information is recorded into homogeneous substrates through predictable physical processes, the resulting data remains physically decoupled from any unique material identity. This lack of inherent randomness makes such systems vulnerable to replication, failing to provide the integrated, hardware-level encryption necessary for absolute security.

Nanodiamonds (NDs) have emerged as a premier material candidate for integrated security hardware (*21, 22*). They are cost-effective, solution-processable, and offer sub-wavelength spatial addressability while retaining the extraordinary properties inherent to the diamond materials, *i.e.*, profound chemical inertness, high thermal stability, and tunable optical properties of internal color centers, such as the nitrogen-vacancy (NV) center (*23, 24*). Crucially, harnessing the idiosyncratic atomic environment of these centers enables a quantum-secured paradigm that provides ultimate physical unclonability and fundamental immunity against emerging algorithmic threats. Despite these advantages, leveraging NDs for high-capacity cryptographic memory remains a formidable challenge due to the lack of deterministic information writing methods at the single-nanoparticle level. Unlike bulk diamond, where femtosecond laser or ion irradiation provides precise spatial control (*25-27*), color-center engineering in single NDs has remained stymied, because existing macroscopic high-energy techniques often induce collateral structural damage or uncontrollable defect yields within the confined, heterogeneous volumes of nanoparticles (*28, 29*). Consequently, ND-based information applications are currently restricted to disordered, read-only authentication tags (*21, 22*), while their vast potential as ordered, programmable cryptographic memory remains untapped.

To overcome these limitations, we establish, for the first time, a deterministic protocol for single-ND color-center customization by systematically exploiting focused-electron-beam irradiation (*30, 31*). By identifying the specific parametric windows that govern vacancy dynamics within ND crystal lattices, we effectively bypass the structural and spatial heterogeneities inherent to the ND ensemble. This approach allows us to tailor the fundamental optical signatures of individual NDs, specifically their emission intensity and spectral profiles, with unprecedented reproducibility and spatial precision. Such capability represents a transformative shift from stochastic ensemble doping to on-demand information encoding at the single-nanoparticle level. Consequently, we provide the first viable pathway for establishing ordered, high-dimensional data storage within randomly dispersed NDs platforms, effectively transforming these stochastically distributed particles into programmable cryptographic units.

Building upon this nanoscale precision, we propose an all-in-one, hardware-intrinsic quantum-secured cryptographic memory (QSCM) architecture (**Fig. 1a**). Our technique enables an ultra-fast single-step customization that independently modulating the absolute photoluminescence (PL) intensity and the spectral ratio of NV to neutral vacancy (GR1) centers in single NDs (**Fig. 1b**). This establishes a multi-dimensional parameter space for high-capacity optical memory. Crucially, this customized data layer is inextricably locked by a dual-key encryption shield native to the stochastic medium. The random spatial coordinates of the dispersed NDs function as a robust macroscopic PUF (Key 1), while inherent strain and electric environment around NV centers (*32, 33*) within each ND yield irreproducible atomic-scale quantum fingerprints (Key 2) that are readily extracted via optically detected magnetic resonance (ODMR). By fusing high-dimensional data encoding with unclonable quantum-spin signatures, our system ensures that stored information is physically inseparable from its unique encryption keys. This hardware-intrinsic approach eliminates the vulnerabilities of decoupled systems, establishing a definitive solid-state paradigm for quantum-secure memory chips, tamper-proof zero-trust supply chains, and next-generation highly resilient data storage.

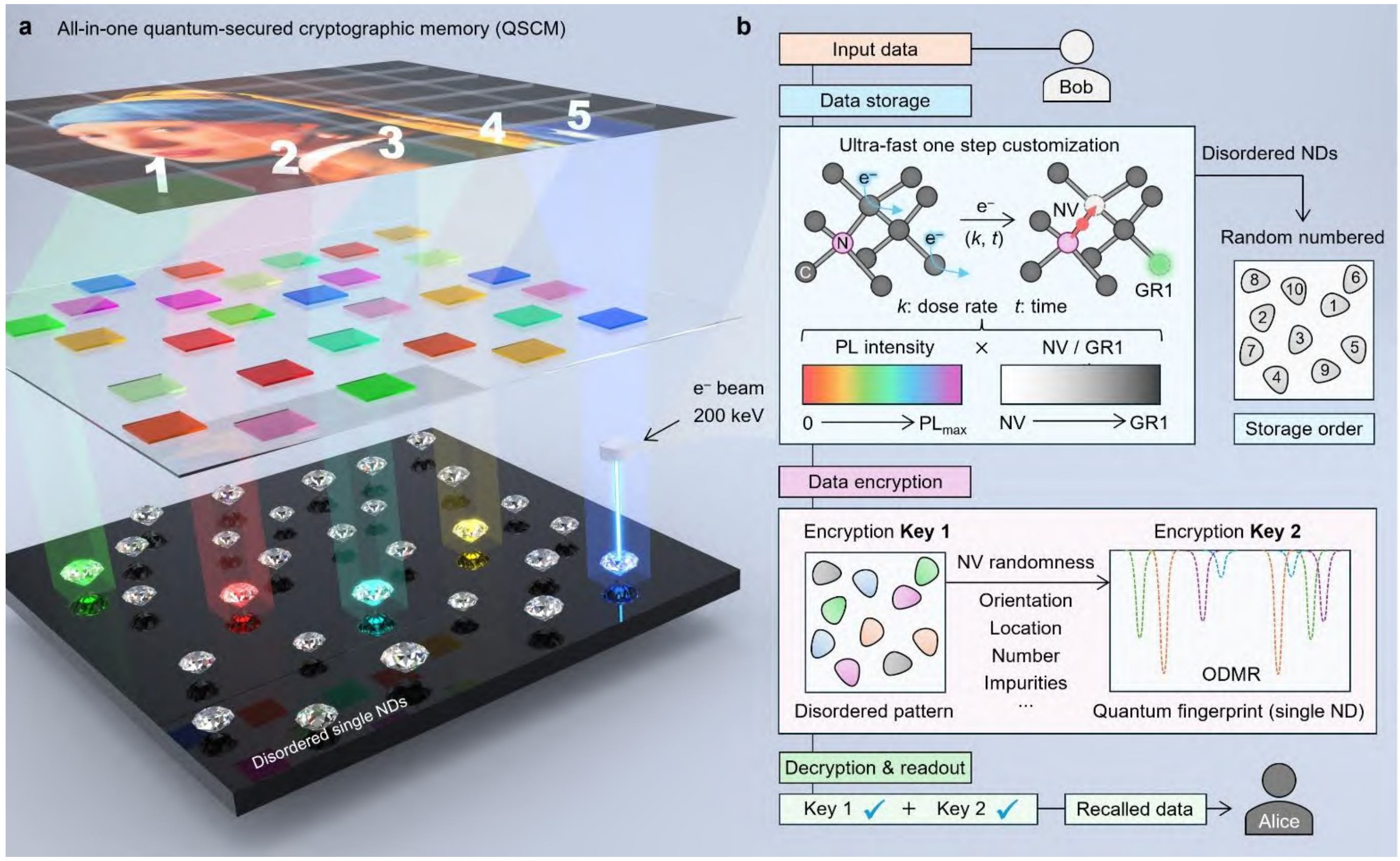

**Fig. 1: All-in-one hardware-intrinsic QSCM using customizable single NDs. a,** Schematic illustrating the targeted customization of individual NDs within a disordered region using a focused electron beam. The technique deterministically encodes high-dimensional data (*e.g.*, pixel data of the artwork *Girl with a Pearl Earring*, from Mauritshuis, The Hague) into single NDs. **b**, System flowchart of the integrated storage and encryption architecture. Data is encoded by Bob via an ultra-fast one-step customization that modulates PL intensity and the NV/GR1 ratio by tuning irradiation dose rate ($k$) and time ($t$). The stored data is intrinsically locked by a dual-key mechanism: Key 1 utilizes the disordered spatial pattern of the NDs as a macroscopic PUF, and Key 2 relies on the irreproducible ODMR quantum fingerprint of each customized ND. Ultimately, Alice can successfully decrypt and retrieve the original data only through the simultaneous verification of both keys.

## Results

### Ultrafast deterministic color-center engineering at the single-nanodiamond level

The realization of ND-based cryptographic memory necessitates a fundamental transition from stochastic ensemble treatments to deterministic and high-throughput manipulation at the single-nanoparticle level. We achieved this by repurposing the focused-electron-beam within a commercially available transmission electron microscope (TEM) from a passive imaging tool into a high-precision writing head (*30*) that locally drives defect evolution inside a selected single ND while preserving nanoscale spatial targeting. During the correlative TEM-optical characterization of NDs (milled from Type Ib high-pressure high-temperature bulk diamond), we serendipitously unveiled an unexpected optical dynamic: contrary to the established studies where the electron beam was generally assumed to have no impact (*33-35*), we discovered that the PL intensity of individual nanoparticles is systematically and deterministically modulated by electron exposure (**Fig. 2**).

As shown in **Fig. 2a**, the PL intensity (cps) of individual NDs follows a strictly deterministic decay as a function of irradiation time (s) across various dose rates (modulated via TEM magnification, kx, see **Table S2**), indicating a highly controllable defect evolution process. Time-lapse TEM imaging further confirmed that, within the irradiation window used here, the overall morphology of the targeted ND remained unchanged during beam exposure (**Fig. S1**). In the context of data storage, this PL controllability allows us to precisely program specific brightness levels as discrete encoding states within a sub-second regime (~1 s), providing a foundation for high-speed optical data writing. Furthermore, the fidelity of information readout depends heavily on the optical quality and contrast of the encoded states. Spectral analysis of a customized single ND (**Fig. 2b**) reveals distinct emission signatures, dominated by a prominent negatively charged nitrogen-vacancy ($NV^-$) zero-phonon line (ZPL) at 637 nm. The remarkable contrast (30.20%) and narrow full-width at half-maximum (FWHM, 2.01 nm) confirm the generation of high-quality NV centers, ensuring high signal-to-noise ratios for reliable data retrieval.

To demonstrate the spatial precision and reproducibility of this protocol, we engineered a spatially ordered NDs bright spots array with programmed brightness levels (**Fig. 2c**). The resulting PL mapping and corresponding intensity profiles confirm that the optical state of each spot can be independently addressed and reliably programmed without detectable cross-talk from adjacent spot. Benchmarking this approach against state-of-the-art bulk and powder diamond NV generation methods (*36-43*) highlights a transformative gain in engineering efficiency (**Fig. 2d, Table S3**). While traditional macroscopic treatments necessitate hours of broad-beam exposure followed by prolonged high-temperature annealing (*37, 40*) to achieve optimal NV concentrations, our localized protocol attains comparable yields through a single-step, ultrafast (seconds-scale) irradiation process, entirely bypassing the need for any post-processing. This dramatic throughput enhancement, representing a leap from hours to seconds, is fundamentally driven by the concentrated electron flux and the highly confined interaction volume within the sub-wavelength nanoparticles. By circumventing both the long-term irradiation and the secondary thermal treatment required by conventional techniques, our one-step methodology effectively surmounts the writing-speed bottleneck inherent in solid-state atomic memory, providing a high-efficiency and scalable pathway for the deployment of quantum-secure information hardware.

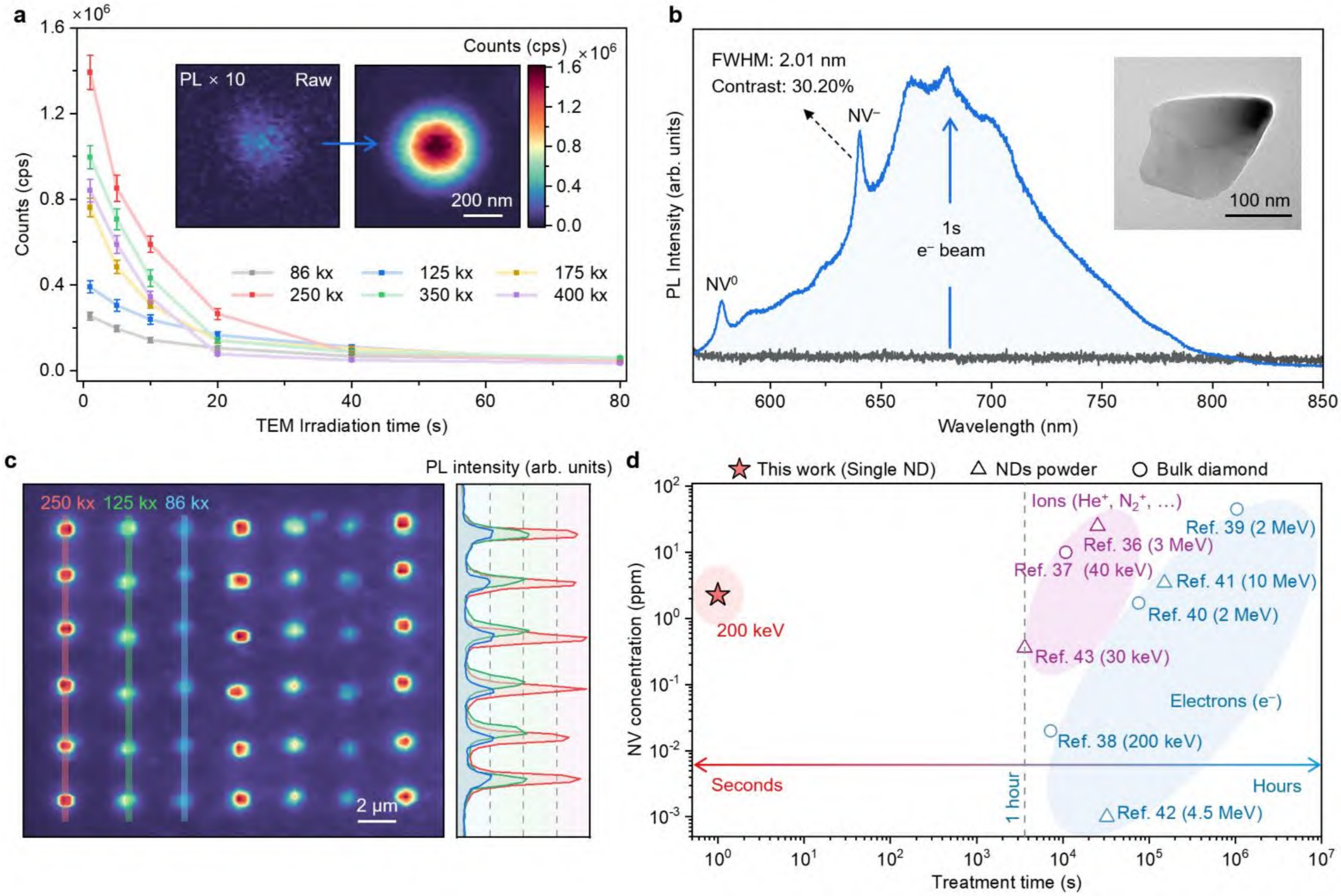

**Fig. 2: Deterministic single-ND color-center engineering via focused-electron-beam irradiation. a,** PL intensity decay of individual NDs as a function of electron beam irradiation time across various dose rates. Insets: Representative PL images of a targeted ND before (left) and after (right) irradiation, demonstrating precise spatial localization. **b**, Representative PL spectrum of a single ND after 1 s (250 kx) of electron-beam irradiation, revealing distinct NV signatures. Inset: Corresponding TEM image of the customized ND. **c**, PL intensity mapping (left) and corresponding spatial intensity profiles (right) of an engineered NDs bright spots array by our approach, demonstrating independent, cross-talk-free brightness modulation at different dose rates. **d**, Benchmarking of our single-ND protocol against existing diamond NV generation methods. Our approach achieves optimal NV concentrations within seconds via a one-step process, significantly outperforming conventional bulk/powder treatments that require hours of exposure and mandatory thermal annealing.

**Dynamic evolution of spectral profiles and quantum fingerprints**

While the deterministic control demonstrated above establishes PL intensity as a programmable writing axis, a secure, high-capacity architecture requires at least one additional independent physical degree of freedom. A defining observation in our study is the progressive reduction in absolute PL intensity during extended electron exposure, a phenomenon that is highly counterintuitive for typical color center generation (*38*). To understand the nature of this anomalous trend, we investigated the defect evolution pathways in NDs under electron beam irradiation and identified a deterministic sequential transition across three distinct optical states (**Fig. 3a**). This multidimensional evolution is fundamentally governed by the competitive kinetics between nitrogen-vacancy coupling and vacancy accumulation within the confined ND crystal lattice.

Initially (~1 s), the lattice undergoes localized vacancy generation adjacent to substitutional nitrogen atoms, actively promoting rapid NV formation (State 1). This is indicated by an NV-dominant spectrum with characteristic zero-phonon lines (blue trace in **Fig. 3d**). As irradiation persists, the local consumption of available nitrogen causes newly generated vacancies (GR1 centers, ZPL at 741 nm) to aggregate, creating a mixed NV/GR1 regime (State 2, purple traces in **Fig. 3d**). Prolonged irradiation eventually culminates in a GR1-dominant state (State 3, green trace in **Fig. 3d**). Importantly, these accumulating GR1 centers do not merely alter the emission spectrum; they act as effective non-radiative quenching sites that suppress NV radiative efficiency, thereby explaining the concurrent reduction in overall PL intensity (**Fig. 2a**).

To rigorously validate these mechanisms, we further developed a coupled kinetic model that effectively maps the electron-beam-driven defect evolution to the optical readout (details in **Materials and Methods**). The model captures the dynamic competition between the electron-beam-assisted activation of NV centers and the progressive generation of vacancy-related defects that drive both the spectral shift and the NV quenching process (*44*). By applying the developed vacancy-induced NV quenching model and performing a global fit across all dose-rate datasets, we successfully reproduced the experimental trajectories (**Fig. 3c**). This analysis confirms that the NV / (NV + GR1) spectral ratio is effectively disentangled from the PL intensity, establishing a second physical axis that exponentially expands the data representation capacity.

Crucially, this defect-evolution process converts the internal quantum environment of each ND into a particle-specific physical unclonable key. Under progressive electron irradiation, the localized structural evolution induces complex lattice strain fields and modifies the local electric environment (*45-47*), which directly perturbs the NV spin Hamiltonian (*47*) (**Fig. 3b**):

$$H = (D + \Pi_z)S_z^2 + (\delta B_z + A_{zz}I_z)S_z + \Pi_z\left(S_y^2 - S_x^2\right) + \Pi_z\left(S_xS_y + S_yS_x\right) \quad (1)$$

This perturbation is physically manifested as idiosyncratic shifting $\Pi_z$ and splitting $2\Pi_\perp$ ($\Pi_\perp = \sqrt{\Pi_x^2 + \Pi_y^2}$) of the $|m_s = \pm 1\rangle$ spin resonances in the corresponding ODMR spectra (**Fig. 3e**). Because the resulting local perturbation is inextricably dictated by the unique, stochastic pristine geometry and impurity distribution of each individual ND, the electron-induced perturbation locks a particle-specific ODMR signature into the ND. Consequently, our writing protocol produces a deterministic, high-dimensional data layer (via PL intensity and spectral ratio) while simultaneously imprinting an atomic quantum fingerprint that is materially inseparable from the stored information.

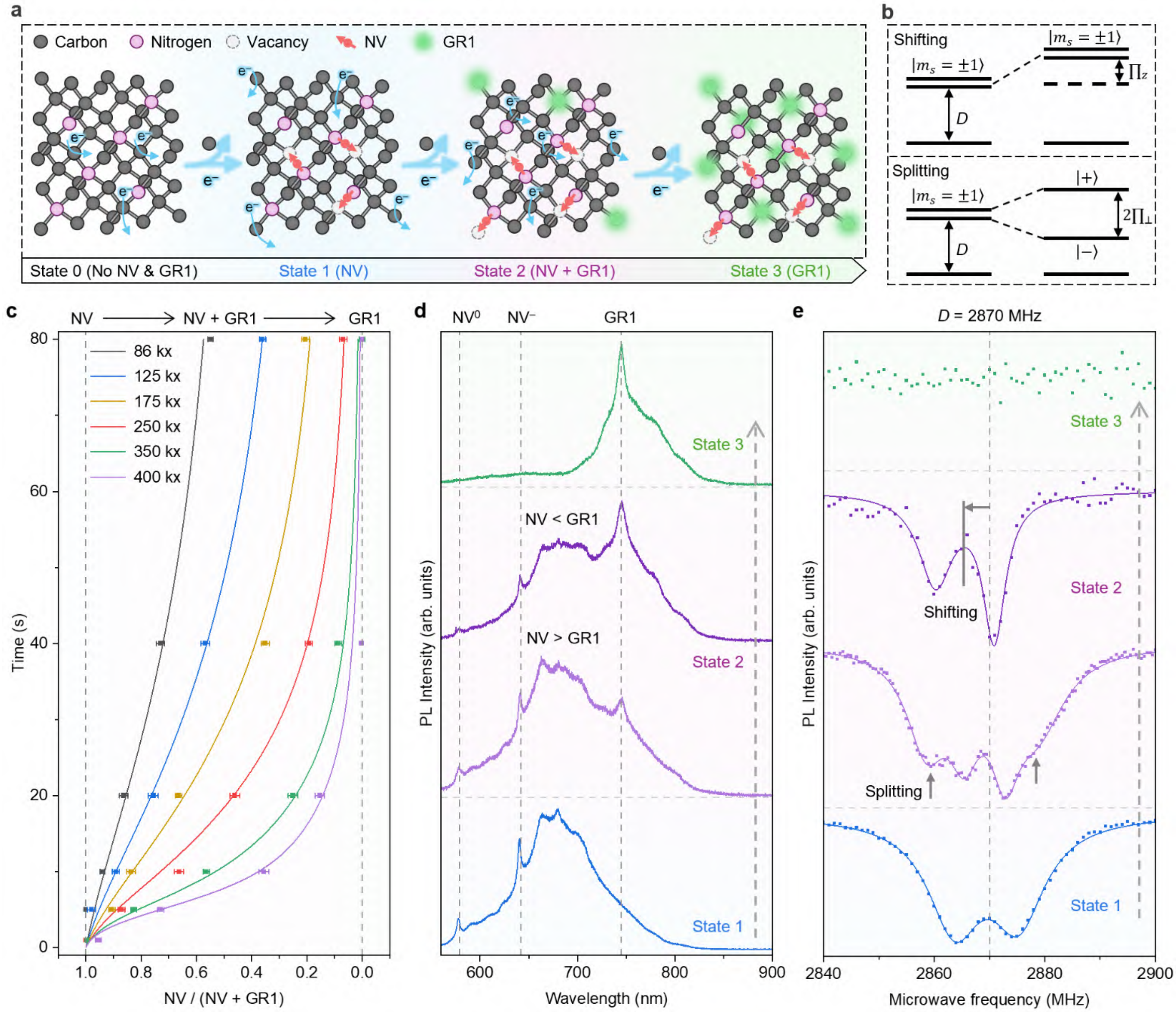

**Fig. 3: Physical mechanism of dynamic spectral evolution and quantum fingerprinting.** **a**, Atomic-scale schematic of the defect evolution pathway during electron beam irradiation, transitioning from an intact lattice (State 0) to an NV-dominant state (State 1), an NV/GR1 mixed state (State 2), and a GR1-dominant state (State 3). **b**, Energy level diagrams illustrating the zero-field splitting ($D$) and shifting of the NV center $|m_s = \pm 1\rangle$ ground states, governed by local lattice strain. **c**, Precise tunability of the NV / (NV + GR1) spectral ratio as a function of irradiation time and dose rate. **d**, Continuous evolution of the PL spectra, confirming the deterministic transition across different optical states. **e**, Corresponding evolution of the ODMR spectra. The unique shifting and splitting of the resonant dips at different states act as an unclonable microscopic quantum fingerprint native to each ND's local lattice environment.

**Deterministic high-dimensional encoding in single nanodiamonds**

The concurrent control over absolute PL intensity and the spectral ratio effectively establishes two independent physical encoding axes, transitioning NDs from simple binary tags to multidimensional information carriers. To harness this capacity for high-density storage, we systematically mapped the defect-evolution kinetics across a two-dimensional (2D) parameter space (**Fig. 4a**). By calibrating the focused electron flux, specifically through the combinatorial pairing of dose rate ($k$) and irradiation time ($t$), we can deterministically navigate this space to program specific pairs of PL intensity and spectral profiles. By leveraging these independent physical degrees of freedom, this multidimensional nanoscale tunability significantly enhances the information entropy of individual NDs, establishing a robust foundation for high-density data multiplexing at the single-nanoparticle level

The robustness of an optical memory relies on the clear distinguishability of its encoded states. As plotted in the experimental 2D phase map (**Fig. 4b**), we successfully isolated 36 distinct optical states within individual NDs. All the PL spectra acquired across the dose-time matrix is provided in **Fig. S2**. The tight clustering of data points and minimal experimental variance (indicated by the narrow error bars in **Fig. 4b**) confirm that these states are separated by distinct boundaries. This base-36 encoding effectively compresses $\log_2(36) \approx 5.17$ bits of information into a single nanoparticle, providing high physical entropy that simultaneously maximizes data density and prevents cloning by exponentially increasing hardware-level complexity. The selection of a 36-state system (Base-36) is highly strategic: it perfectly accommodates the universally utilized alphanumeric character set (10 digits, 0–9; and 26 letters, A–Z), enabling the direct and intuitive encoding of complex human-readable strings without the overhead of binary translation. Furthermore, the inherent precision of our customization technique theoretically permits the resolution of even higher-dimensional states, for instance, by implementing finer gradations in electron dosimetry to narrow the state boundaries, or by integrating additional optical degrees of freedom such as fluorescence lifetime (*48*). Nevertheless, the current 36-state discretization provides an optimal engineering balance between ultra-high information density and zero-error optical readout.

To validate the scalability and practical writing fidelity of this high-dimensional encoding architecture, we performed a proof-of-concept data archiving experiment. We addressed six individual NDs within a disordered array to encode the string "NATURE" (**Fig. 4c**). By applying the predetermined dose-time writing matrix, the optical states of the six targeted NDs were successfully driven to their designated coordinates in the codebook. Optical readout demonstrated a perfect one-to-one correspondence between the measured PL intensities, the extracted spectral ratios, and the intended alphanumeric characters. This 100% decoding fidelity underscores the precision of our localized electron-beam protocol and confirms the viability of using single NDs as programmable, ultra-high-density optical memory units, fully prepared for integration with the underlying quantum-spin encryption layer.

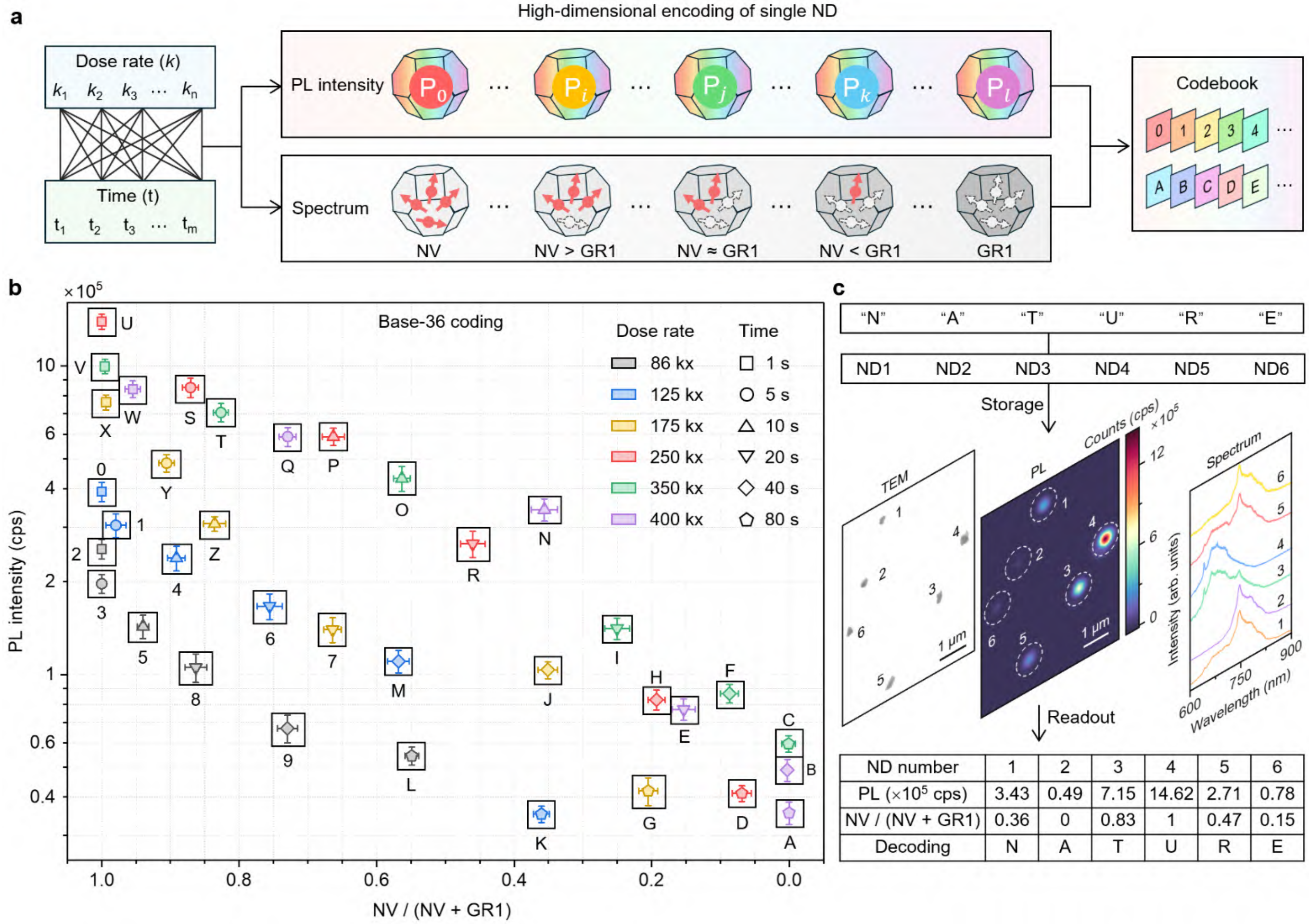

**Fig. 4: Deterministic high-dimensional data encoding at the single-nanoparticle level. a,** Schematic of the high-dimensional encoding strategy. Combinatorial inputs of focused-electron-beam dose rate ($k$) and irradiation time ($t$) allow the PL intensity and spectral profiles to be decoupled and mapped to a high-dimensional codebook. **b**, Experimental 2D phase map demonstrating 36 distinct, resolvable optical states within individual NDs. The well-separated coordinates along the PL intensity and NV / (NV + GR1) axes yield a high-dimensional Base-36 alphanumeric encoding (0–9, A–Z) in single NDs, providing an optimal engineering balance between data density and readout fidelity. **c**, Proof-of-concept demonstration of deterministic data writing and storage. Six spatially distinct single NDs (TEM image, middle left) were customized to encode the string "NATURE". The optical readout PL mapping (middle center) and extracted individual spectra (middle right) show successful retrieval of the intended characters based on the Base-36 codebook (table, bottom).

## Quantum-secured cryptographic memory architecture

The fusion of deterministic high-dimensional encoding with the intrinsic stochasticity of NDs enables an all-in-one QSCM architecture, where information is physically inseparable from its unique encryption keys. To demonstrate this, we conceptualized an integrated quantum cryptographic micro-ID chip with customizable functional modules (**Fig. 5a**). This versatile architecture allows for the partitioned storage of diverse data types, including alphanumeric basic information, pixel-level imagery, and dedicated backup areas reserved for subsequent data writing. Security within this chip is enforced by a dual-key multiscale framework. The primary layer (Key 1) leverages the macroscopic spatial PUF provided by the disordered NDs. To demonstrate this, the alphanumeric string "GIRLWITHAPEARLEARRING" was encoded into randomly selected NDs within a specific region (**Fig. 5b**). Because the encoded characters are spatially dispersed, their precise coordinates and readout sequence constitute a high-entropy permutation key. As shown in **Fig. 5c**, attempting to read this array using an incorrect sequence triggers a permutation trap, yielding only cryptographic noise (*e.g.*, "AWALK..."), whereas the correct spatial key flawlessly decrypts the stored data.

The ultimate security of this chip is guaranteed by the atomic quantum fingerprints (Key 2), extracted via ODMR. We characterized the spin signatures of 20 NDs selected directly from the encoded NDs in **Fig. 5b**, observing extreme heterogeneity in their resonance frequencies and splitting patterns (**Fig. 5d**). This variance arises from the idiosyncratic atomic-scale strain and electric fields environment within each ND, which are permanently locked by the electron-beam customization process. A cross-correlation analysis of the decryption fidelity yields a high-fidelity diagonal matrix (**Fig. 5e**), confirming that no two NDs within the storage medium share the same quantum signature.

To validate the high-fidelity archival capability of this system for complex datasets, *e.g.*, visual information, we performed the end-to-end storage and retrieval of a full-color digital image (**Fig. 5f**). The process involved an initial analysis of the target image's pixel values, which were then mapped onto our discrete base-36 codebook. Each color profile was translated into a specific deterministic optical state and written into a randomly dispersed ND (**Fig. 5g**). While an incorrect storage order results in an unintelligible pixel matrix, the simultaneous application of the spatial PUF and ODMR quantum fingerprints allows for the 100% fidelity reconstruction of the *Girl with a Pearl Earring* portrait (**Fig. 5h**), as the deterministic Base-36 state-mapping provides a unique and unambiguous decoding solution. This integrated hardware-intrinsic architecture fundamentally resolves the long-standing vulnerability of decoupled hardware: the information is not merely algorithmically encrypted, but is materially embedded within an irreproducible quantum environment, providing an unconditionally secure foundation for hardware-level data safeguarding.

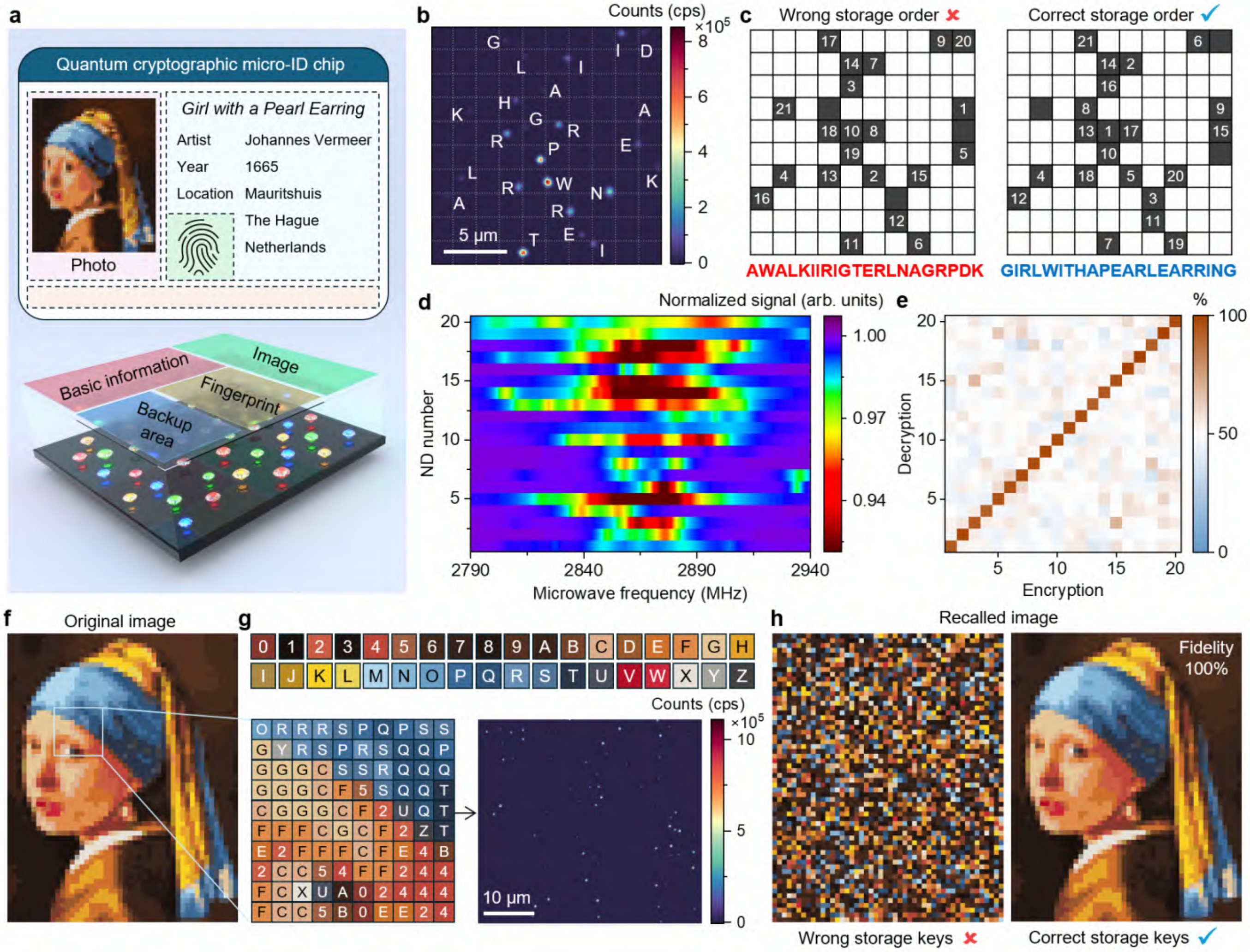

**Fig. 5: Implementation of the all-in-one QSCM architecture. a,** Conceptual design of an integrated quantum cryptographic micro-ID chip featuring customizable modules for alphanumeric information, imagery, and backup areas for future expansion, *etc*. **b**, PL mapping of randomly dispersed NDs serving as a macroscopic spatial PUF, wherein specific NDs are customized to encode the string "GIRLWITHAPEARLEARRING". **c**, Demonstration of the spatial encryption mechanism (Key 1). Decoding with a randomized (wrong) storage order yields unintelligible noise, whereas the correct coordinate matrix flawlessly retrieves the encrypted string. **d**, ODMR spectra map for 20 NDs selected from (**b**), highlighting the extreme diversity in their quantum fingerprints (Key 2). **e**, Decryption fidelity matrix confirming the unclonability of the individual ND quantum signatures. **f–h**, End-to-end encode and recall of the painting *Girl with a Pearl Earring* from Mauritshuis, The Hague. The original image (**f**) is analyzed and mapped to the base-36 codebook for deterministic writing into disordered single NDs (**g**) and successfully retrieved with 100% fidelity using the correct encryption keys, whereas applying wrong keys results in failed decryption (**h**).

**System-level protocol for hardware-intrinsic zero-trust architecture**

The inextricable fusion of deterministic high-dimensional storage and quantum-level unclonability within single NDs establishes a formidable foundation for hardware-intrinsic security. To illustrate its real-world utility, we propose a system-level validation protocol utilizing this integrated architecture to provide hardware-intrinsic security within a zero-trust framework (**Fig. 6**). In this paradigm, sensitive data is first encoded and archived within the integrated NDs chip. During validation, the data server enforces a multi-tiered authentication protocol that strictly requires the simultaneous verification of both the macroscopic spatial sequence (Key 1) and the microscopic atomic quantum fingerprints (Key 2). Attack attempts, including direct hardware substitution (fake chip A) or cloning via replication (fake chip B), are intrinsically rejected. This multi-layered robustness is guaranteed by the irreproducible properties of our NDs, precluding cloning even by adversaries with equivalent fabrication capabilities.

By providing a fully integrated, *in-situ* logic within a microscopic form factor, our platform reduces the risk of data leakage and ensures that information is materially inseparable from its encryption. The architecture leverages quantum-level randomness to raise the barrier against machine-learning modelling and side-channel attacks, while simultaneously offering the high-capacity programmable memory required for complex data archival. These combined capabilities, ultrafast writing kinetics, high information entropy, and multiscale encryption, establish a performance benchmark against existing storage and encryption technologies (**Table S1**). Beyond its cryptographic superiority, the exceptional chemical and thermal stability of diamond ensures that the stored information and its associated encryption keys remain resilient under harsh environments (*18, 21*). By resolving the long-standing vulnerability of decoupled memory and security components, this work provides an unconditionally secure solid-state foundation for quantum-resistant authentication and high-fidelity archival data vaults.

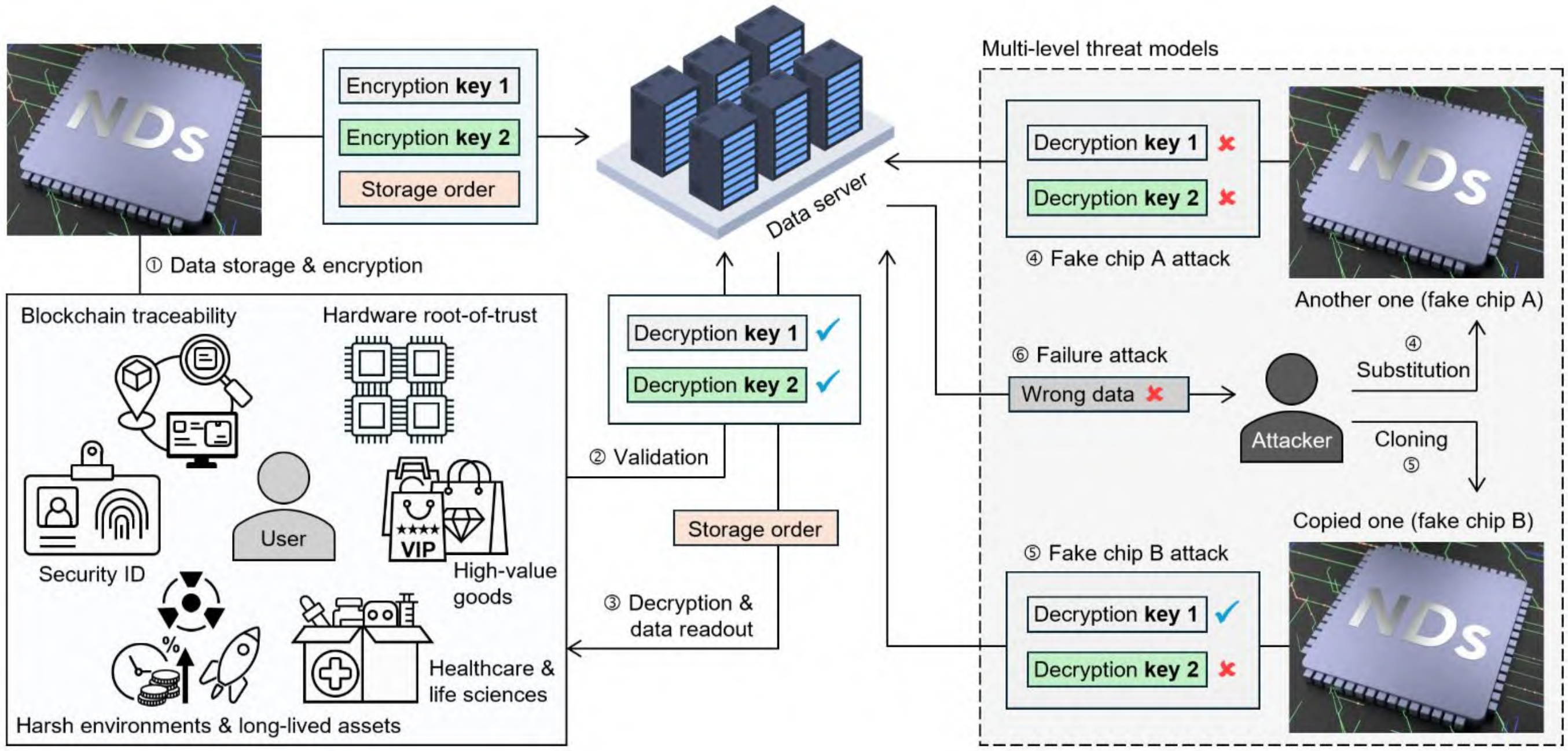

**Fig. 6: System-level hardware-intrinsic security framework and practical application scenarios.** Schematic flowchart illustrating the end-to-end validation protocol and real-world utility of the integrated NDs Chip. ① Data storage and encryption: High-dimensional data is deterministically written and materially locked within the NDs chip. ②–③ Validation and readout: The data server enforces a multi-tiered authentication process. Successful decryption and data retrieval require the simultaneous verification

of the macroscopic spatial PUF (Key 1) and the microscopic atomic quantum fingerprints (Key 2). ④–⑥ Rejection of attackers: The architecture intrinsically rejects unauthorized access attempts. The substitution attack (fake chip A) fails due to an incorrect spatial coordinate sequence (Key 1), whereas the cloning attempt (fake chip B), even if manufactured using the identical writing protocol, is rejected owing to the mismatch in its idiosyncratic quantum fingerprints (Key 2). This platform's unique combination of high security and material robustness provides immediate deployment pathways for diverse high-stakes applications, including hardware root-of-trust, blockchain traceability, security identification, high-value asset protection, and long-term data archiving in harsh environments.

## Discussion

In summary, we have established an all-in-one hardware-intrinsic cryptographic memory architecture that inextricably fuses high-dimensional data storage with multiscale quantum encryption at the single-nanoparticle limit. By repurposing focused-electron-beam irradiation into a deterministic nanoscale writing tool, we overcame the long-standing stochasticity of diamond nanocrystal defect engineering. This one-step, ultrafast protocol achieves the concurrent modulation of PL intensity and spectral profiles, realizing an unprecedented base-36 data encoding capacity within individual NDs. Crucially, this programmable data layer is materially locked by the intrinsic macroscopic spatial randomness and the irreproducible microscopic atomic-spin fingerprints of single NDs. This all-in-one architecture fundamentally eliminates the vulnerability gap inherent in conventional systems where memory and cryptographic components are physically decoupled.

Looking forward, the scalability of this single-nanoparticle writing protocol and its compatibility with established optical readout techniques present a highly viable pathway for industrial deployment. As sophisticated cyber-threats, machine-learning modelling, and quantum computing advance, the global demand for unconditionally secure, hardware-intrinsic cryptography will only intensify. By transforming disordered NDs ensembles into programmable, tamper-proof cryptographic units, this work provides a robust physical foundation for next-generation zero-trust architectures. Coupled with the exceptional resilience of diamond in extreme environments, our platform paves the way for ultra-secure hardware roots-of-trust, tamper-evident global supply chains, and quantum-resistant archival data vaults, marking a definitive leap forward in solid-state information technology.

## Materials and Methods

### Deterministic customization of single-nanodiamond

Following our established protocol (*49*), salt-assisted air-oxidized nanodiamonds (SAAO NDs) were utilized as the raw storage medium. Briefly, a mixture comprising 10 mg of HPHT NDs (200 nm, PolyQolor) and 50 mg of sodium chloride (NaCl, 99.5%, Sigma-Aldrich) was air-oxidized at 500 °C for 2 hours. The NaCl residue was subsequently removed via centrifugation with deionized (DI) water. The diluted SAAO ND suspension (0.1 mg/mL) was drop-cast onto a silicon dioxide TEM window grid (Electron Microscopy Sciences) to ensure the target NDs were well-dispersed as single, isolated entities. For deterministic data writing, the TEM grid was loaded into a transmission electron microscope (FEI, Tecnai G2 20 S-TWIN). The focused-electron-beam operated at an accelerating voltage of 200 keV. Target NDs were initially located using low-magnification imaging (<5kx). To execute the writing process, the electron dose rate ($k$) was modulated by increasing the TEM magnification to the designated level (ranging from 86 to 400 kx) and maintained for a precisely defined irradiation time ($t$), followed by rapid beam blanking/magnification reduction. This one-step, localized electron-beam irradiation was executed without any subsequent thermal annealing, ensuring a rapid, *in-situ* defect customization. The NV concentration in **Fig. 2d** was estimated by normalizing its PL intensity to the PL intensity of reference ND containing single NV center (**Fig. S3**). The concentration (in ppm) of NV centers was estimated based on the ratio of the number of NV centers to the total number of carbon atoms within the ND volume. For this estimation, each ND was approximated as a rounded, disk-like shape, thicker in the center and tapering towards the ends, with its height being one-third of its diameter (*50*).

**Scanning confocal fluorescence readout and spectral decoupling analysis**

PL intensity readout and spatial mapping were performed using a custom-built scanning confocal microscope equipped with a nanopositioning stage (Physik Instrumente, P-562.3CD). Excitation was provided by a 532 nm continuous-wave laser (10 μW), delivered through a single-mode fiber with circular polarization (>95%), and focused on the NDs using an air objective (Olympus, LMPlanFLN, 100X/0.8). Fluorescence emission was collected via single-mode fused fiber optic couplers (Thorlabs, TW670R5F1) and detected by single-photon counting modules (Excelitas Technologies, SPCM-AQRH-16-FC). For spectral decoding, the collected fluorescence was transmitted to a spectrometer (Teledyne Princeton Instruments, FERGIE-ISO-81) with a 550 nm long-pass filter (LBTEK, MEFH10-550LP) isolating the excitation laser. The optical state of each ND was quantified by two parameters: absolute PL intensity and the spectral ratio $f = A_{NV} / (A_{NV} + A_{GR1})$. Here, $A_{NV}$ represents the integrated PL area assigned to the NV emission, and $A_{GR1}$ is the integrated area of the GR1 center (details see **Supplementary Text 1**). The NV ZPL characteristics (FWHM and contrast) were extracted using a Lorentzian-exponential fitted equation (*51*).

**Kinetic modelling of defect evolution during electron-beam irradiation**

To quantitatively validate the defect evolution mechanism, we established a coupled kinetic model that maps the electron-beam-driven defect dynamics to the spectral ratio $f(t)$. The model employs a system of flux-dependent rate equations to capture the dynamic competition among NV center activation and the progressive accumulation of GR1. To correlate these underlying state populations with the optical readout, a vacancy-induced exponential quenching formalism was applied. A global fit was subsequently performed simultaneously across all experimental dose-rate datasets using shared kinetic parameters, successfully reproducing the theoretical trajectories. Full mathematical formulations, definitions of state variables, and detailed modelling procedures are provided in **Supplementary Text 2**.

**Extraction of ODMR quantum fingerprints**

To extract the quantum fingerprints of the customized NDs, ODMR measurements were conducted using continuous microwaves. The microwave signals were generated by a signal generator (Rohde & Schwarz, SMIQ03B), amplified by a 45 dB amplifier (Mini-Circuits, ZHL-16W-43-S+), and delivered to a custom microwave structure situated near the ND sample. The localized accumulation of GR1 defects during electron-beam customization significantly amplifies the internal lattice strain and local charge heterogeneity within each ND. This amplification exacerbates the idiosyncratic shifting and splitting of the NV ground-state $|m_s = \pm 1\rangle$ spin resonances, vastly increasing the inter-particle variance and rendering these optical states highly robust as physical fingerprints. ODMR spectra were acquired by sweeping the microwave frequency and recording the PL intensity response under constant experimental parameters (laser power: 50 μW; microwave power: −15 dBm). To rigorously define and compare the quantum signatures across different NDs, each ODMR spectrum was transformed into a compact digital fingerprint that captures particle-specific spectral features. Pairwise similarity between different NDs was then evaluated based on these fingerprints, as detailed in **Supplementary Text 3**.

**High-dimensional data encoding and image archiving protocol**

The Base-36 codebook was established by calibrating the continuous 2D physical parameter space mapped by the electron-beam dose rate and irradiation time. Thirty-six distinct optical states, defined by unique pairs of PL intensity and spectral ratio, were assigned to an alphanumeric character set (0–9, A–Z). For text encoding, character sequences were translated into specific writing matrices (*k*, *t*) and sequentially programmed into pre-selected single NDs. For image archiving, the original digital portrait was first pixelated; the color depth of each pixel was digitally mapped to the closest discrete state within our Base-36 codebook. The corresponding writing matrix was then executed on randomly dispersed NDs. During decryption, the system cross-referenced the macroscopic spatial readout sequence (Key 1) and verified the ODMR signatures (Key 2) before rendering the PL and spectral data back into the intended visual or textual format.

**Acknowledgements**
Z.C. acknowledges the financial support from the National Natural Science Foundation of China (NSFC) and the Research Grants Council (RGC) of the Hong Kong Joint Research Scheme (Project No. N_HKU750/23), the Shenzhen-Hong Kong-Macau Technology Research Programme (Category C project, no. SGDX20230821091501008).

**Author contributions**
Z.C. supervised the project. T.Z. and Z.C. conceived the ideas and designed the experiments. T.Z. led the experimental investigation. T.Z. and J.L. performed the TEM irradiation. J.L. and T.Z. performed the optical measurements. T.Z. J.L., and Z.C. wrote the manuscript.

**Competing interests**
The authors declare no competing interests.

**Data and materials availability**
All data are in the main text or the supplementary materials.